\documentstyle[12pt,epsfig]{article}
\def\beqa{\begin{eqnarray}}
\def\eeqa{\end{eqnarray}}
\def\beq{\begin{equation}}
\def\eeq{\end{equation}}

\def\half{\frac{1}{2}}

\def\half{\frac{1}{2}}

\def\l{{\cal L}}

\def \phan{\phantom}

\def\pr{{\it Phys. Rev.}\ }
\def\prl{{\it Phys. Rev. Lett.}\ }
\def\pl{{\it Phys. Lett.}\ }

\def\ijmp{{\it Int. Journ. Mod. Phys.}\ }

\def\cqg{{\it Class. Quantum Grav.}\ }
\def\aph{{\it Ann. Phys. (Leipzig)}\ }

\def\apj{{\it Ap. J.}\ }
\def\aj{{\it Astron. J.}\ }
\def\aa{{\it Astron. Astrophys.}\ }
\def\ncim{{\it Nuovo Cim.}\ }

\def\rmp{{\it Rev. Mod. Phys.}\ }

\def\ass{{\it Astr. and Space Sci.}\ }

\begin{document}
\begin{titlepage}
        \title{Matching torsion $\Lambda-$term with observations}

\author{S. Capozziello$^{(1),(2)}$\thanks{capozziello@sa.infn.it},
V.F. Cardone$^{(1),(2)}$\thanks{winny@na.infn.it}, E.
Piedipalumbo$^{(2),(3)}$\thanks{ester@na.infn.it},\\ M.
Sereno$^{(2),(3)}$\thanks{sereno@na.infn.it},
A. Troisi$^{(1),(2)}$\thanks{antro@sa.infn.it}\\
 {\em\small $^{(1)}$Dip. di Fisica "E.R. Caianiello", Universit\'a di Salerno} \\
 {\em\small Via S. Allende, I-84081 Baronissi (Sa), Italy.}\\
 {\em \small$^{(2)}$Istituto Nazionale di Fisica Nucleare, Sez. di Napoli,} \\
 {\em \small $^{(3)}$Dip. di Scienze Fisiche, Universit\'a  di Napoli "Federico II"}\\
 {\em\small Monte S. Angelo, Via Cinthia, Edificio N I-80126 Napoli, Italy.}\\}

\date{\today}

\maketitle

\begin{abstract}
Taking into account a torsion field gives rise to a negative
pressure contribution  in cosmological dynamics and then  to an
accelerated behaviour of  Hubble fluid.   The presence of torsion
has the same effect of a $\Lambda$-term. We  obtain a general
exact solution which well fits  data coming from high redshift
supernovae and Sunyaev-Zeldovich/X-ray method for the
determination of cosmological parameters. On the other hand, it
is possible to obtain observational constraints on the amount of
torsion density. A dust dominated Friedmann behaviour is
recovered as soon as torsion effects are not relevant.
\end{abstract}

\thispagestyle{empty} \vspace{20.mm}
 PACS number(s): 98.80.Cq, 98.80. Hw, 04.20.Jb, 04.50 \\

\vspace{5.mm}

\vfill

\end{titlepage}

\section{Introduction}

A generalization  of Einstein General Relativity can be obtained
by considering a torsion tensor different from zero in ${\bf U}_4$
space-time manifold where connection is not symmetric
\cite{hehl,trautman}. Such an approach is very interesting today
in relation to several extended theories of gravity as
Superstrings, Supergravity and Kaluza-Klein theories. In
particular, torsion allows to include spin matter fields in
General Relativity and the Einstein-Cartan-Sciama-Kibble (ECKS)
theory is one of most serious attempt in this direction. However,
not all the forms of torsion are directly connected to a spin
counterpart as it is widely discussed in \cite{classtor}.

Now if some forms of torsion allow to take into account spin in
General Relativity, it seems reasonable that they could have had
some role into dynamics of the early universe when   the number
density of particles per volume was huge. The presence of torsion
gives naturally a repulsive contribution to the energy-momentum
tensor \cite{desabbata}. In fact, it is possible to show that for
densities of the order of $10^{47} g/cm^3$ for electrons and
$10^{54} g/cm^3$ for protons and neutrons, torsion could give
observable consequences if all the spins of the particles result
aligned. These huge densities can be reached only in the early
universe so that cosmology is the only viable approach to test
torsion effects \cite{desabbata}. However no relevant tests
confirming the presence of torsion have been found until now and
it is still an open debate if the space-time is a Riemannian
${\bf V}_4$ manifold or not. Considering cosmology and, in
particular primordial phase transitions and inflation
\cite{desabbata,kolb,peebles}, it seems very likely that, in some
regions of early universe, the presence of local magnetic fields
could have aligned  the spins of particles. At very high
densities, this effect could influence the evolution of
primordial perturbations remaining as an imprint in today
observed large scale structures.

>From another point of view, the presence of torsion could give
observable effects without taking into account clustered matter
but resulting as a sort of cosmological constant.
Recent observations  seem to point out  that
the universe is accelerating.
 Type Ia Supernovae (SNe Ia)
\cite{perlmutter},  data coming from clusters of galaxies
\cite{cluster} and
 CMBR investigations \cite{boomerang} give
observational constraints from which we deduce that the universe is
spatially flat, low density and dominated by some kind of
non-clustered dark energy. Such an energy, which is supposed to
have dynamics, should be the origin of the observed cosmic acceleration.

In terms of density parameter, we have
$\Omega_m\simeq 0.3\,,\Omega_{\Lambda}\simeq 0.7\,, \Omega_{k}\simeq 0.0\,,$
where $\Omega_{m}$ includes  non-relativistic baryonic and
non-baryonic (dark) matter, $\Lambda$ is the dark energy
(cosmological constant, quintessence,..), $k$ is the curvature
parameter of Friedmann-Robertson-Walker (FRW) metric.

Standard matter fluid as source of Einstein--Friedmann
cosmological equations gives rise to expanding decelerated
dynamics. To fit observations, non-standard forms of
matter-energy have to be taken into account: the net effect
should be to implement a sort of a {\it cosmological constant}
which naturally gives rise to a negative pressure capable of
implementing an accelerated cosmic expansion. On the other hand,
{\it Quintessence} \cite{steinhardt,rubano,curvature} generalizes
this approach taking into account all the mechanisms which give
rise to negative pressure regimes for cosmic fluid. In particular,
scalar fields.

The cosmological constant problem is one of the main issue of
modern physics since its value  should provide the
gravity vacuum state \cite{weinberg}, should be connected  to
the mechanism which led the early universe to the today observed
large scale structures \cite{guth,linde}, and should predict
what will be the fate of the whole universe (no--hair conjecture)
\cite{hoyle}.

In any case, we need a time variation of cosmological constant to
get successful inflationary models, to be in agreement with
 observations,
and to obtain a  de Sitter stage
 toward the future. In other words, this means that cosmological
constant has to acquire a great value in early epoch, it has to
undergo a phase transition with a graceful exit
and has to result in a small remnant toward the future coinciding with the
observational constraints {\it (coincidence problem)}.

The today observed  accelerated cosmological
behaviour should be the result of this dynamical process where
the present  value of cosmological constant is not fixed exactly at  zero.

In this context, a fundamental issue is to select the classes of
gravitational theories and  conditions which "naturally" allow
to recover an effective cosmological constant without considering
special initial data. Theories with torsion could match up this point,
as we will see below, also if the coincidence problem (that is the fine
tuning between huge initial values of $\Lambda$ and very thin today observed
constraints) remain essentially unsolved.

This paper is organized as follows. Sec.2 is devoted to an
essential summary of gravity with torsion. In Sec.3, we show how
introducing torsion in cosmological dynamics gives rise to a
negative pressure extra-term acting as a cosmological constant
and we found a general cosmological solution. In Sec.4, we  match
such a cosmological solution  with data coming from SNe Ia surveys
and Sunyaev-Zeldovich/X-ray method. Sec.5 is devoted to discussion
and conclusions.

\section{Gravity with torsion}
In this section, we give general definitions of torsion and
associated quantities. We shall use the notation in \cite{hehl}
and in \cite{classtor}. It is a convention to call {\bf U}$_4$ a
$4$-dimensional space-time manifold endowed with metric and
torsion. The manifolds with metric and without torsion are
labeled as {\bf V}$_4$.

Torsion tensor $S_{ab}^{\phan{ab}c}$ can be defined, in a ${\bf U}_4$
space-time manifold, using the antisymmetric part
of the affine connection coefficients $\Gamma_{ab}^{c}$, that is
 \beq
 \label{t1}
S_{ab}^{\phan{ab}c}=\frac{1}{2}\left(\Gamma_{ab}^{c}-\Gamma_{ba}^{c}\right)
\equiv\Gamma_{[ab]}^{c}\,, \eeq where $a,b,c = 0,\dots 3$.

In Einstein General Relativity, it is postulated that
$S_{ab}^{\phan{ab}c}=0$.

Often in the calculations, torsion occurs in linear combinations
as in the {\it contortion tensor}, defined as
 \beq
\label{t3}
K_{ab}^{\phantom{ab}c}=-S_{ab}^{\phantom{ab}c}-S^{c}_{\phan{c}a b}
+
S^{\phantom{a}c}_{b\phantom{c}{a}}=-K^{\phantom{a}c}_{a\phan{c}b}\,,
\eeq

and in the {\it modified torsion tensor}

\beq \label{t4} T_{ab}^{\phantom{ab}c}=
S_{ab}^{\phantom{ab}c}+2\delta^{\phan{[a}c}_{[a}S_{b]}
 \eeq
 where $S_a\equiv S_{ab}^{\phantom{ab}b}$.

According to these definitions, it follows that the affine
connection can be written as \beq \label{t5}
\Gamma_{ab}^{c}=\left\{^{c}_{ab}\right\}-K_{ab}^{\phantom{ab}c}\,,
\eeq where $\left\{^{c}_{ab}\right\}$ are the usual Christoffel
symbols of the symmetric connection.

The torsion  contributions to the Riemann tensor  can be
explicitly given by
\begin{equation}\label{riexpanded}
R_{abc}^{\phantom{abd}d} =R_{abc}^{\phantom{abd}d}(\{\}) -
 \nabla_{a}K_{bc}^{\phantom{b]c}d} +  \nabla_{b}K_{ac}^{\phantom{ac}d}
+ K_{ae}^{\phantom{ae}d}K_{bc}^{\phantom{bc}e}-
K_{be}^{\phantom{be}d}K_{ac}^{\phantom{ac}e}
\end{equation}
where $R_{abc}^{\phantom{abc}d}(\{\})$ is the standard Riemann tensor
of General Relativity.

>From Eq.(\ref{riexpanded}), the expressions for the Ricci tensor
and the curvature scalar are
\begin{equation}\label{ricci}
 R_{ab}= R_{ab}(\{\}) - 2\nabla_{a}S_{c} + \nabla_{b}K_{ac}^{\phantom{ac}b}
+ K_{ae}^{\phantom{ae}b}K_{bc}^{\phantom{bc}e}-
2S_eK_{ac}^{\phantom{ac}e}
\end{equation}
and
\begin{equation}\label{curvscalar}
 R=R(\{\}) - 4 \nabla_{a}S^{a} + K_{ceb}K^{bce} - 4 S_aS^a.
\end{equation}

The closest theory to General Relativity containing torsion is the
Einstein-Cartan-Sciama-Kibble (ECSK) theory. It is described by

\beq \label{t6}
 L=\sqrt{-g}\left(\frac{R}{2k}+{\l}_{m}\right),
\eeq which is  the Lagrangian density of General Relativity
depending on the metric tensor $g_{ab}$ and on the connection
$\Gamma_{ab}^{c}$. $R$ is the curvature scalar
(\ref{curvscalar}) and ${\l}_{m}$ the Lagrangian density of
matter fields.

By the variation of the matter Lagrangian density with respect to the metric
we get

 \beq \label{t7}
\mbox{{\bf t}}^{ab}=\frac{\delta {\cal L}_{m}}{\delta g_{ab}},
 \eeq
which is the symmetric stress--energy tensor while the variation with respect
to the contortion tensor
 \beq \label{t8}
\tau_{c}^{\phan{c}ba}=\frac{\delta {\cal L}_{m}}{\delta
K_{ab}^{\phan{ab}c}},
 \eeq
is the source of torsion. In many instances, it can be identified
with a spin density, but there are many cases in which
the source of the torsion
field is not spin \cite{classtor}.

By the full variation of (\ref{t6}) and introducing the canonical
energy-momentum tensor
\begin{equation}\label{canonical}
{\bf \Sigma^{ab}}= {\bf t^{ab}} + {}^*\tilde\nabla_{c}(\tau^{abc}
- \tau^{bca} +\tau^{cab})\, ,
\end{equation}
where we used the abridged notation
${}^*\tilde\nabla_{c}:=\tilde\nabla_{c} +2S_{cd}^{\phantom{cd}d}$,
the following field equations are derived \cite{hehl}
\begin{equation}\label{t9}
   G^{ab}=8\pi G {\bf\Sigma}^{ab}\, ,
\end{equation}
and
 \beq \label{t10}
 T_{ab}^{\phan{ab}c}=8\pi G\tau_{ab}^{\phan{ab}c}\,,
\eeq
 where
$c=1$, and the symbol $\tilde\nabla$ indicates covariant
derivative with torsion.

Eqs.(\ref{t9})  generalize the Einstein equations in a
$U_4$.

 Eqs.(\ref{t10}) are algebraic so that it is
always possible to cast Eqs.(\ref{t9}) in  pure Einstein ones, by
substituting the torsion terms with their sources. The result is the
definition of an effective energy--momentum tensor as the source of the
Riemannian part of the Einstein tensor \cite{hehl}. In doing so,
one obtains

\beq \label{t11}
  G^{ab}(\{\})=8\pi G\tilde{\mbox{{\bf t}}}^{ab},
  \eeq

where $G^{ab}(\{\})$ is the Riemannian part of the Einstein
tensor. The effective energy--momentum tensor is then

 \beqa
\label{t12} \tilde{\mbox{{\bf t}}}^{ab}&=&\mbox{{\bf t}}^{ab}+
8\pi G\left[-4\tau^{ac}_{\phantom{ac}[d}
\tau^{bd}_{\phantom{bd}c]}-
2\tau^{acd}\tau^{b}_{cd}+\tau^{cda}\tau_{cd}^{\phantom{cd}b}\right.\nonumber\\
&+&\left.\half
g^{ab}(4\tau^{\phantom{e}c}_{e\phantom{c}[d}\tau^{ed}_{\phantom{ed}c]}
+\tau^{ecd}\tau_{ecd})\right].
 \eeqa

The tensor ${\bf t}^{ab}$ can be of the  form

 \beq \label{t13} \mbox{{\bf
t}}^{ab}=(p+\rho)u^{a}u^{b}-pg^{ab} , \eeq

if standard perfect--fluid matter is considered.

These considerations are extremely useful for the following,
since torsion contributions can be treated as additional
perfect-fluid terms.

\section{Cosmology with a torsion $\Lambda$-term}

The totally antisymmetric part of torsion
 can be expressed by a 4-vector \beq
\sigma^a=\epsilon^{abcd}S_{bcd}\,. \eeq If one imposes to it the
symmetries of a background which is homogeneous and isotropic, it
follows that, in comoving coordinates, only the component
$\sigma^0$ survives as a function depending only on cosmic time
(see \cite{classtor,goenner,tsamparlis}).

For a perfect fluid, the Einstein-Friedmann cosmological
equations can be written as

\beq \frac{\ddot{a}}{a}=-\frac{4\pi G
}{3}(\tilde{\rho}+3\tilde{p})\,, \eeq

and

 \beq \left(\frac{\dot{a}}{a}\right)^{2}+\frac{k}{a^{2}}=
\frac{8\pi G}{3}\tilde{\rho}\,, \eeq

where $a(t)$ is the cosmological scale factor and $k=0,\pm 1$ is
the spatial curvature constant. Energy density and pressure can
be assumed in the forms
 \cite{classtor,goenner,tsamparlis}

\beq \label{density} \tilde\rho=\rho +  f^2\,, \qquad
\tilde p= p -f^2\,, \eeq

where $f$ is a function related to $\sigma^0$, while $\rho$ and $p$ are
the usual quantities of General Relativity. This choice can be
pursued since
  we can define $S_{abc}=S_{[abc]}=f(t)$ where $f(t)$ is a
 generic function of time which we consider as the source of torsion.
 For a detailed discussion of this point see \cite{classtor}

As above, we can define a stress-energy tensor of the form \beq
\mbox{{\bf
t}}^{(tot)}_{ab}=(\tilde{p}+\tilde{\rho})u_{a}u_{b}-\tilde{p}g_{ab}\,,
\eeq which, by Eqs. (\ref{density}), can be splitted as \beq
\mbox{{\bf t}}^{(tot)}_{ab}=\mbox{{\bf
t}}^{(matter)}_{ab}+\mbox{{\bf t}}^{(torsion)}_{ab}\,. \eeq Due to
the contracted Bianchi identity, we have \beq \label{bianchi}
\mbox{{\bf t}}^{(tot);b}_{ab}=0\;; \eeq from which, we can assume
that (cfr. \cite{minkowski}) \beq \label{bianchi1}
 \mbox{{\bf t}}^{(matter);b}_{ab}=0\,,\qquad \mbox{{\bf t}}^{(torsion);b}_{ab}=0\,. \eeq In the FRW space-time,
(\ref{bianchi}) becomes
 \beq
\dot{\tilde{\rho}}+3H(\tilde\rho + \tilde p)=0 \eeq which is \beq
\label{continuity} \dot{\rho} + 3H(\rho +  p) = -2f\dot{f}\,. \eeq
>From Eqs.(\ref{bianchi1}), both sides of (\ref{continuity}) vanish
independently so that \beq f(t)=f_{0}=\mbox{constant.} \eeq In
other worlds, a torsion field gives rise to a constant energy
density. Taking into account standard matter whose equation of
state is defined into the above Zeldovich range, we obtain

\beq \label{dense}
 \tilde{\rho}=\rho_{0}\left[\frac{a_0}{a}
\right]^{3(\gamma+1)}+f_{0}^2\,. \eeq

Inserting this result into the cosmological equations, we obtain,
in any case,  a monotonic expansion being $f_{0}^2>0$,
$\dot{a}^2>0$. We obtain an  accelerated behaviour if \beq
\label{acceleration} \rho+3p<2f_{0}^2\,, \eeq so that
acceleration depends on the torsion density.

In a dust-dominated universe, we have \beq \label{dust}
\tilde{\rho}=\rho_{0}\left(\frac{a_{0}}{a}\right)^{3}+f_{0}^{2}\,,
\qquad \tilde{p}=-f_{0}^2\,,\eeq and then the general solution is
\beq \label{general}
a(t)={\left(\frac{a_0^3\rho_0}{2f_0}\right)^{1/3}
\left[\cosh(f_{0}t)-1\right]}^{1/3}\,. \eeq

Obviously, if $f_{0}t\rightarrow 0$, we have $\displaystyle{\cosh
(f_{0}t)\simeq (f_{0}t)^2}$ and then $a\sim t^{2/3}$, as it has to
be.

This results tell us that a $\Lambda$-term can be obtained without
considering additional scalar fields in the dynamics but only
assuming that the space-time is ${\bf U}_4$ (a manifold with
torsion) instead of ${\bf V}_4$ (the standard Riemann manifold of
General Relativity). The advantage to pass from ${\bf V}_4$ to
${\bf U}_4$ is due to the fact that the spin of a particle turns
out to be related to the torsion just as its mass is responsible
for the curvature. From this point of view, such a generalization
includes the spin fields of matter into the same geometrical
scheme of General Relativity \cite{hehl}.

In what follows, we will compare the obtained solution with data
coming from observations.

\section{ Matching  torsion $\Lambda$-term with observations}

\subsection{The Supernovae SNe Ia method}

It is well known that the use of astrophysical standard candles
provides a fundamental mean of measuring the cosmological
parameters. Type Ia supernovae (SNe\,Ia) are the best candidates
for this aim since they can be accurately calibrated and can be
detected at enough high red-shift. This fact allows to
discriminate among cosmological models. To this aim, one can fit
a given model to the observed magnitude\,-\,redshift relation,
conveniently expressed as\,:

\begin{equation}
\mu(z) = 5 \log{\frac{c}{H_0} d_L(z)} + 25 \label{eq: muz}
\end{equation}
being $\mu$ the distance modulus and $d_L(z)$ the dimensionless
luminosity distance. The distance in the model we are considering
is completely equivalent to the one in a spatially flat universe
with a non-zero cosmological constant. Thus $d_L(z)$ is simply
given as\,:

\begin{equation}
d_L(z) = (1 + z) \int_{0}^{z}{dz' [ \Omega_M (1 + z')^3 +
\Omega_{tor} ]^{-1/2}} \ . \label{eq: dl}
\end{equation}
where $\Omega_ {tor} = 1 - \Omega_M$ plays the same role as the
usual $\Omega_{\Lambda}$.

The distance modulus, as said,  can be obtained from observations
of SNe\,Ia. The apparent magnitude $m$ is indeed measured, while
the absolute magnitude $M$ may be deduced from template fitting or
using the Multi\,-\,Color Lightcurve Shape (MLCS) method. The
distance modulus is then simply $\mu = M - m$. Finally, the
redshift $z$ of the supernova can be determined accurately from
the host galaxy spectrum or (with a larger uncertainty) from the
supernova spectrum.

Our model can be fully characterized by two parameters\,: the
today Hubble constant $H_0$ and the matter density $\Omega_M$. We
find their best fit values minimizing the $\chi^2$ defined as\,:

\begin{equation}
\chi^2(H_0, \Omega_M) = \sum_{i}{\frac{[\mu_i^{theor}(z_i | H_0,
\Omega_M) - \mu_i^{obs}]^2} {\sigma_ {\mu_{0},i}^{2} +
\sigma_{mz,i}^{2}}} \label{eq: defchi}
\end{equation}
where the sum is over the data points \cite{wang}. In Eq.(\ref{eq:
defchi}), $\sigma_{\mu_{0}}$ is the estimated error of the
distance modulus and $\sigma_{mz}$ is the dispersion in the
distance modulus due to the uncertainty $\sigma_z$ on the SN
redshift. We have\,:

\begin{equation}
\sigma_{mz} = \frac{5}{\ln 10} \left ( \frac{1}{d_L}
\frac{\partial d_L}{\partial z} \right ) \sigma_z \label{eq:
sigmamz}
\end{equation}
where we assume $\sigma_z = 200 \ \rm{km \ s^{-1}}$ adding in
quadrature $2500 \ \rm{km \ s^{-1}}$ for those SNe whose redshift
is determined from broad features \cite{wang}. Note that
$\sigma_{mz}$ depends on the cosmological parameters so that we
will employ an iterative procedure to find the best fit values.

The High\,-\,z team and the Supernova Cosmology Project have
detected a quite large sample of high redshift ($z \simeq 0.18 -
0.83$) SNe\,Ia, while the Calan\,-\,Tololo survey has investigated
the nearby sources. Using the data in Perlmutter et al.
\cite{perlmutter} and Riess et al. \cite{riess}, we have compiled
a combined sample of 85 SNe as described in detail in \cite{wang}.
We exclude 6 lixely outliers SNe as discussed in
\cite{perlmutter}, thus ending with a total of 79 SNe. The
results of the fit are presented in Fig.1 where we show the 1,2
and 3\,$\sigma$ confidence regions in the $(\Omega_M, H_0)$ plane.

\begin{figure*}[ht]
\centering \resizebox{10cm}{!}{\includegraphics{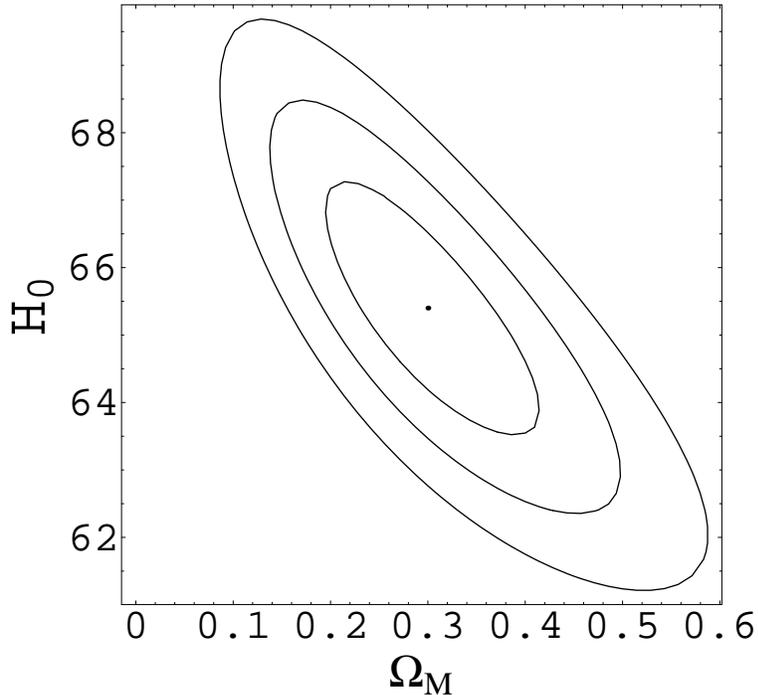}}
\hfill \caption{1, 2 and 3$\sigma$ confidence regions in the
$(\Omega_M, H_0)$ plane. The central dot represents the best fit
values\,: $\Omega_M = 0.3, H_0 = 65.4 \ {\rm km \ s^{-1} \
Mpc^{-1}}$.}
\end{figure*}

The best fit values (with $1 \sigma$ error) turn out to be\,:

\begin{displaymath}
\Omega_M = 0.30 \pm 0.08 \ \ , \ \ H_0 = 65.4 \pm 1.2 \ km \
s^{-1} \ Mpc^{-1} \ .
\end{displaymath}
which allow to conclude that a torsion $\Lambda$-term could
explain observation very well. On the other hand, we can estimate
the torsion density contribution which result  to be\,:

\begin{displaymath}
f_0^2 = (5.6 \pm 0.7) \times 10^{-30} \ g \ cm^{-3} \,,
\end{displaymath}
which is a good value if compared to the cosmological critical
density. It is worthwhile to note that, in the case of SNe, the
error on $H_0$ does not take into account systematic uncertainties
due to possible calibration errors. Due to this reason it is so
small with respect to other results in literature.

\subsection{The Sunyaev-Zeldovich/X-ray method}

Besides the above results,  we can discuss how the Hubble constant
$H_0$ and the torsion density parameter $\Omega_{\it tor}$ can be
constrained also by the angular diameter distance $D_A$ as
measured using the Sunyaev-Zeldovich effect (SZE) and the thermal
bremsstrahlung (X-ray brightness data) for galaxy clusters.
Distances measurements using SZE and X-ray emission from the
intracluster medium are based on the fact that these processes
depend on different combinations of some parameter of the clusters
(see \cite{birk} and references therein). The SZE is a result of
the inverse Compton scattering of the CMB photons of hot electrons
of the intracluster gas. The photon number is preserved, but
photons gain energy and thus   a decrement of the temperature is
generated in the Rayleigh-Jeans part of the black-body spectrum
while an increment rises up in the Wien region. We will limit our
analysis to the so called {\it thermal} or {\it static} SZE,
which is present in all the clusters, neglecting the ${\it
kinematic}$ effect, which is present in those clusters with a
nonzero peculiar velocity with respect to the Hubble flow along
the line of sight. Typically the thermal SZE is an order of
magnitude larger than the kinematic one.
 The shift of temperature is:
\begin{equation}
\frac{\Delta T}{T_0} = y\left[ x \, \mbox{coth}\,
\left(\frac{x}{2} \right) -4 \right], \label{eq:sze5}
\end{equation}
where ${\displaystyle x=\frac{h \nu}{k_B T}}$ is a dimensionless
variable, $T$ is the radiation temperature, and $y$ is the so
called Compton parameter, defined as the optical depth $\tau =
\sigma_T \int n_e dl$ times the energy gain per scattering:
\begin{equation}\label{compt}
  y=\int  \frac{K_B T_e}{m_e c^2} n_e \sigma_T dl.
\end{equation}
In the Eq.~(\ref{compt}), $T_e$ is the temperature of the
electrons in the intracluster gas, $m_e$ is the electron mass,
$n_e$ is the numerical density of the electrons, and $\sigma_T$ is
the  cross section of Thompson electron scattering. We have used
the condition $T_e \gg T$ ($T_e$ is the order of $10^7\, K$ and
$T$, which is the CBR temperature is $\simeq 2.7K$). Considering
the low frequency regime of the Rayleigh-Jeans approximation we
obtain
\begin{equation}
 \frac{\Delta T_{RJ}}{T_0}\simeq -2y
 \label{eq:sze5bis}
\end{equation}
The next step  to quantify the SZE decrement is the one to specify
the models for the intracluster electron density and temperature
distribution. The most commonly used model is the so called
isothermal $\beta$ model  \cite{cavaliere}. We have
\begin{eqnarray}
& & n_e (r) = n_e (r) = n_{e_0} \left( 1 + \left(
\frac{r}{r_e} \right)^2 \right)^{-\frac{3 \beta}{2}} \\
& & T_e (r) = T_{e_0}~, \label{eq:sze6}
\end{eqnarray}
being $n_{e_0}$ and $T_{e_0}$, respectively the central electron
number density and temperature of the intracluster electron gas,
$r_e$ and $\beta$ are fitting parameters connected with the model
  \cite{sarazin}. For the effect of the
cluster modelling see \cite{jetzer}. From (6) we have
\begin{equation}
\frac{\Delta T}{T_0} = -\frac{2 K_B \sigma_T T_{e_0} \,
n_{e_0}}{m_e c^2} \cdot \Sigma \label{eq:sze7}
\end{equation}
being
\begin{equation}
\Sigma = \int^\infty_0 \left( 1 + \left( \frac{r}{r_c}
\right)^2\right)^{-\frac{3 \beta}{2}} dr\,. \label{eq:sze8}
\end{equation}
The integral in Eq.~(\ref{eq:sze8}) is overestimated  since
clusters have a finite radius. The effects of the finite
extension of the cluters are analyzed
in~\cite{jetzer},\cite{cooray98b}.

A simple geometrical argument converts the integral in
Eq.~(\ref{eq:sze8}) in angular form, by introducing the angular
diameter distance, so that

\begin{equation}
\Sigma = \theta_c \left(1 + \left(
\frac{\theta}{\theta_2}\right)^2 \right)^{1/2 - 3 \beta/2}
\sqrt{\pi} \, \frac{\Gamma \left( \frac{3 \beta}{2} -
\frac{1}{2}\right)}{\Gamma \left( \frac{3 \beta}{2} \right)} \,
r_{DR}. \label{eq:sze9}
\end{equation}

In terms of the dimensionless angular diameter distances, $d_A$
(such that $D_A=\displaystyle\frac{c}{H_0} d_A$) we get
\begin{equation}
\frac{\Delta T (\theta)}{T_0} = - \frac{2}{H_0}\frac{\sigma_T K_B
T_{ec}n_{e_0}}{m_e c} \sqrt{\pi}  \frac{\Gamma \left( \frac{3
\beta}{2} \frac{1}{2} \right)}{\Gamma \left(
\frac{3\beta}{2}\right)} \left( 1 - \left(
\frac{\theta}{\theta_2}\right)^2 \right)^{\frac{1}{2} (1 -3
\beta)} d_A, \label{eq:sze10}
\end{equation}
and, consequently, for the central temperature decrement, we get
\begin{equation}\label{eq:sze10bis}
\frac{\Delta T (\theta =0)}{T_0}=- \frac{2}{H_0}\frac{\sigma_T
K_B T_{ec}n_{e_0}}{m_e c} \sqrt{\pi} \frac{\Gamma \left( \frac{3
\beta}{2} \frac{1}{2} \right)}{\Gamma \left(
\frac{3\beta}{2}\right)}\frac{c}{H_0} d_A.
\end{equation}
The factor $\displaystyle \frac{c}{H_0} d_A$  in
Eq.~(\ref{eq:sze10bis}) carrys the dependence on the thermal SZE
on both the cosmological models (through $H_0$ and the Dyer-Roeder
distance $d_A$) and the redshift (through $d_A$). From
Eq.~(\ref{eq:sze10bis}), we also note that the central electron
number density is proportional to the inverse of the angular
diameter distance, when calculated through SZE measurements. This
circumstance allows to determine the distance of cluster, and then
the Hubble constant, by the measurements of its thermal SZE and
its X-ray emission.

This possibility is based on the different power laws, according
to which  the decrement of the temperature in the SZE,
$\frac{\Delta T(\theta =0)}{T_0}$ , and the X-ray emissivity,
$S_X$, scale with respect to the electron density. In fact, as
above pointed out, the electron density, when calculated from SZE
data, scales as $d^{-1}_{A}$~( $n^{SZE}_{e0}\propto d^{-1}_A$),
while the same one scales as $d^{-2}_{A}$~($n^{X-ray}_{e0}\propto
d^{-2}_A$) when calculated from X-ray data. Actually, for the
X-ray surface brightness, $S_X$,  assuming for the temperature
distribution of  $T_e=T_{e0}$, we get the following formula:
\begin{equation}\label{sx}
 S_X=\frac{\epsilon_X}{4\pi}{n_{e0}^2}\frac{1}{ (1+z)^3} \theta_c \frac{c}{H_0} d_A I_{SX},
\end{equation}
being \[I_{Sx}=\int^{\infty}_0 \left(\frac{n_e}{n_{e0}}\right)^2
dl\] the X-ray structure integral, and $\epsilon_X$ the spectral
emissivity of the gas (which, for $T_e\geq 3{\times}10^{7}$, can
be approximated by a typical value: $\epsilon_X = \epsilon
\sqrt{T_e}$, , with $\epsilon \simeq 3.0 {\times} 10^{-27} n_p^2$
erg $cm^{-3}$ $s^{-1}$ $K^{-1}$~\cite{sarazin}) . The angular
diameter distance can be deduced by eliminating the electron
density from Eqs.~(\ref{eq:sze10bis}) and (\ref{sx}), yielding:

\begin{equation}
\frac{y^2}{S_X}= \frac{4 \pi (1+z)^3}{\epsilon} {\times}
 \left(\displaystyle\frac{k_B \sigma_{T}}{m_e c^2}\right)^2 {T_{e0}}^{3/2}
 \theta_c \frac{c}{H_0} d_A {\times}
\frac{\left[B(\frac{3}{2}\beta-\frac{1}{2},
\frac{1}{2})\right]^2}{B(3\beta-\frac{1}{2}, \frac{1}{2})}
\label{eq:sze11}
\end{equation}
where $B(a,b)=\displaystyle\frac{\Gamma(a)\Gamma(b)}{\Gamma(a+b)}$
is the Beta function.

It turns out that
\begin{equation}\label{eq:sze11bis}
D_A=\frac{c}{H_0}d_A\propto \frac{(\Delta T_0)^2}{S_{X0}
T^2_{e0}}\frac{1}{\theta_c},
\end{equation}
where all these quantities are evaluated along the line of sight
towards the center of the cluster (subscript 0), and $\theta_c$ is
referred to a characteristic scale of the cluster along the line
of sight. It is evident that the specific meaning of this  scale
depends on the density model adopted for  clusters. In our
calculations we are using the so called $\beta$ model.

Eqs.~(\ref{eq:sze11}) allows  to  compute the Hubble constant
$H_0$, once the redshift of the cluster is known and the other
cosmological parameters are, in same way, constrained. Since the
dimensionless Dyer-Roeder distance, $d_A$,  depends on
$\Omega_{{\it tor}}$, $\Omega_{m}$, comparing the  estimated
values with the theoretical formulas for $D_A$, it is possible to
obtain information about $\Omega_m$ $\Omega_{{\it tor}}$, and
$H_0$. Recently distances of 18 clusters with redshift ranging
from $z\sim 0.14$ to $z\sim 0.78$ have been determined from a
likelihood joint analysis of SZE and X-ray observation (see Table
7 \cite{reese} and reference therein). Modeling the intracluster
gas as a spherical isothermal $\beta $-model allows to obtain
constraints on the Hubble constant $H_0$ in a standard
$\Lambda$-FRW model. We perform a similar analysis using angular
diameter distances measurements for a sample of 44 clusters,
constituted by the 18 above quoted  clusters and other 24 already
know data (see \cite{birk}). In Fig~(2),  our data are plotted
against two theoretical models.

\begin{figure*}[ht]
\label{fig:sample} \centering
\resizebox{10cm}{!}{\includegraphics{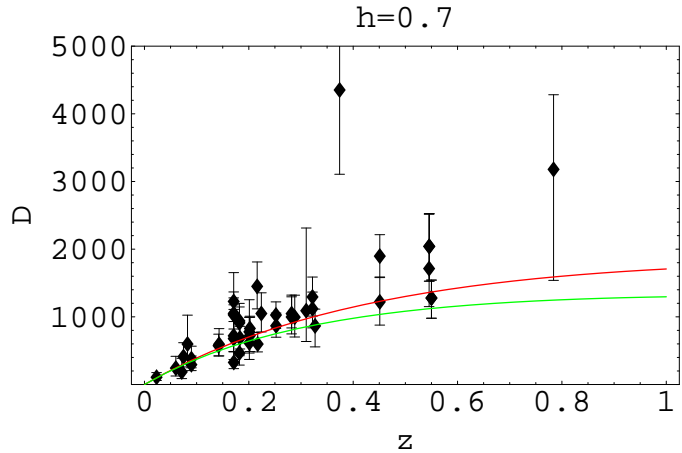}} \hfill
\caption{We plot the measure angular diameter distances with the
error bars and two theoretical model:
 a Einstein- De Sitter model (bottom curve) and a flat $\Lambda$-FRW model with $\Omega_{m}=0.3$,
 $\Omega_{\Lambda}=0.7$ (upper curve), both with $H_0 = 70 \ {\rm km \ s^{-1} \
 Mpc^{-1}}$}
 \end{figure*}
 As indicated in
\cite{ birk, reese},  the errors $\sigma$ are  only of
statistical nature. Taking into account our  model with torsion
(\ref{general}), the theoretical expression for the angular
diameter distances $D_A$ is
\begin{equation}\label{eq:szetors1}
  D_A(z)=\frac{1}{(1+z)^2}d_L(z) = \frac{1}{1+z} \int_{0}^{z}{dz' [ \Omega_M (1 +
z')^3 + \Omega_{tor} ]^{-1/2}}
\end{equation}
We find the best fit values for $ \Omega_{tor}$ and $H_0$,
minimizing the reduced  $\chi^2$:
\begin{equation}
\chi^2(H_0, \Omega_M) = \sum_{i}{\frac{[{D_A}^{theor}(z_i | H_0,
\Omega_M) - {D_A}_i^{obs}]^2}{\sigma_
{D_A}^{2}}}\label{eq:eq:szetors2}
\end{equation}
The results are shown in the Fig.~(3), where we see the contours
corresponding to the $68.5\%$ and $98\%$ confidence levels:
\begin{figure*}[ht]
\label{fig:szetor} \centering
\resizebox{10cm}{!}{\includegraphics{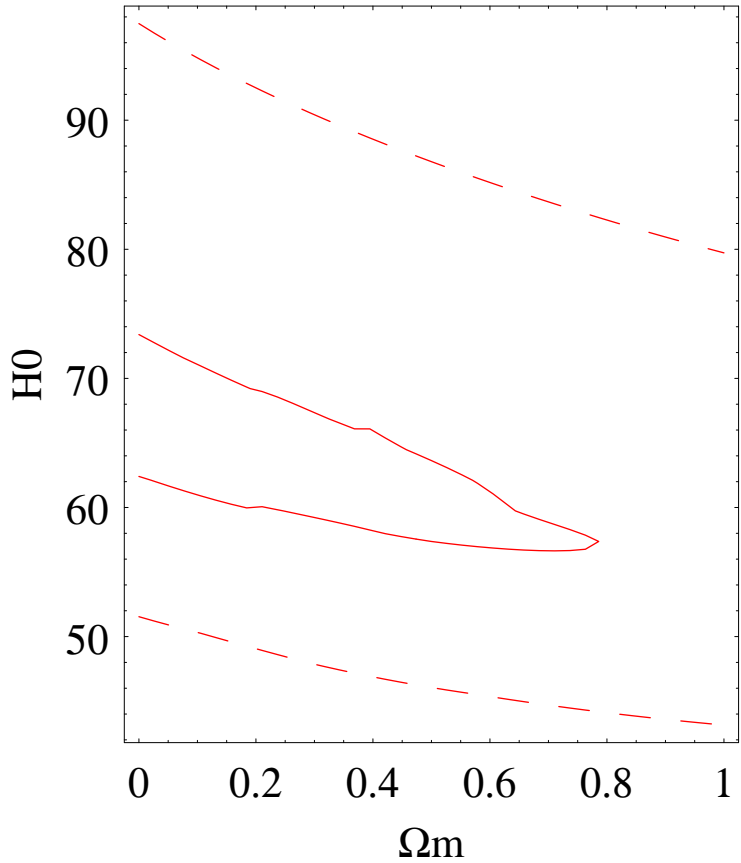}} \hfill
\caption{1, $68.5\%$ and $98\%$  confidence regions in the
$(\Omega_M, H_0)$ plane. The best fit values are\,: $\Omega_M =
0.3, H_0 = 68 \ {\rm km \ s^{-1} \ Mpc^{-1}}$.}
\end{figure*}
The best fit values (at $1 \sigma$) turn out to be\,:
\begin{displaymath}
\Omega_M = 0.30 \pm 0.3 \ \ , \ \ H_0 = 68\pm 6 \ km \ s^{-1} \
Mpc^{-1} \ .
\end{displaymath}

in good agreement with the above fit derived from SNe Ia data.

\section{\normalsize\bf Discussion and conclusions}

In this paper, we dealt with a cosmological model where torsion is
present into dynamics. After the discussion of how such a
contribution modifies cosmological Friedmann-Einstein equations,
we have shown that the net effect of torsion is the introduction
of an extra-term into fluid matter density and pressure which is
capable of giving rise to an accelerated behaviour of the cosmic
fluid. Being such a term a constant, we can consider it a sort of
torsion $\Lambda$-term. If the standard fluid matter is dust, we
can exactly solve dynamics which is in agreement with the usual
Friedmann model (to be precise Einstein-de Sitter) as soon as
torsion contribution approaches to zero.

The next step is to compare the result with observations in order
to see if such a torsion cosmology gives rise to a coherent
picture. We have used SNe Ia data, Sunyaev-Zeldovich effect and
X-ray emission from galaxy clusters. Using our model, we are
capable to reproduce the best fit values of $H_0$ and $\Omega_M$
which gives a cosmological model dominated by a cosmological
$\Lambda$-term. In other words, it seems that introducing torsion
(and then spins) in dynamics allows to explain in a {\it natural}
way the presence of cosmological constant or  a generic form of
dark energy without the introduction of exotic scalar fields.
Besides, as we have seen in Sec.4, observations allows to estimate
torsion density which can be comparable to other forms of matter
energy $(\sim 5.5\times 10^{-30}\,g\,cm^{-3})$.

However, we have to say that we used only a particular form of
torsion and the argument can be more general if extended to all
the forms of torsion \cite{classtor}. Furthermore, being in our
case the torsion contribution a constant density, it is not
possible to solve {\it coincidence} and {\it fine tuning}
problems. To address these issues we need a form of torsion
evolving with time. This will be the topic of a forthcoming paper.


\begin{thebibliography}{99}
\bibitem{hehl}
F.W. Hehl, P. von der Heyde, G.D. Kerlick and J.M. Nester,
\rmp {\bf 48} (1976) 393.
\bibitem{trautman}
A. Trautman {\it Nature} {\bf 242} (1973) 7.
\bibitem{classtor}
S. Capozziello, G. Lambiase, and C. Stornaiolo \aph {\bf 10}
(2001) 8, 713.
\bibitem{desabbata}
V. de Sabbata, \ncim {\bf A 107} (1994) 363.\\
V. de Sabbata and C. Sivaram, \ass {\bf 165} (1990) 51;\\
V. de Sabbata and C. Sivaram, \ass {\bf 176} (1991) 141.
\bibitem{kolb}
E.W. Kolb, and M.S. Turner {\it The Early Universe} \\
Addison-Wesley 1990 (Redwood City, Calif.)
\bibitem{peebles}
P.J.E. Peebles {\it Principle of Physical Cosmology},\\
Princeton Univ. Press 1993 (Princeton).
\bibitem{perlmutter}  B.P. Schmidt et al. \apj {\bf 507}, 46
    (1998).\\ S.  Perlmutter et al.  \apj {\bf 483}, 565 (1997).
    \\ S. Perlmutter et al. {\it Nature} {\bf 391}, 51
    (1998).\\ S. Perlmutter et al. \apj {\bf 517}, 565 (1999).
\bibitem{cluster} B. Chaboyer et al. \apj {\bf 494}, 96 (1998).\\
                  M.Salaris and A. Weiss \aa {\bf 335}, 943 (1998).
\bibitem{boomerang} P. de Bernardis et al. {\it Nature} {\bf 404}, 955 (2000).
\bibitem{steinhardt} R.R. Caldwell,  R. Dave, P.J. Steinhardt,
    \prl {\bf 80}, 1582 (1998).
\bibitem{rubano} R. de Ritis et al., \pr {\bf D 62} (2000) 043506.\\
                 C. Rubano and J.D. Barrow, \pr {\bf D64} (2001)
                 127301.\\
                 C. Rubano and P. Scudellaro, astro-ph/0103335
                 (2001).
\bibitem{curvature} S. Capozziello, \ijmp {\bf D 11}(2002) 483.
\bibitem{weinberg} S. Weinberg, \rmp {\bf 61} (1989) 1.
\bibitem{guth}
A. Guth, \pr  {\bf D 23} (1981) 347;   \pl  {\bf 108 B} (1982)
389.
\bibitem{linde} A.D. Linde, \pl {\bf B 108}, 389 (1982);
         \pl {\bf B 114}, 431 (1982);
          \pl {\bf B 238} (1990) 160.
\bibitem{hoyle} F. Hoyle and J.V. Narlikar, {\it Proc. R. Soc.} {\bf 273A} (1963) 1.
\bibitem{goenner} H. Goenner and F. M\"uller-Hoissen, \cqg {\bf 1} (1984) 651.
\bibitem{tsamparlis} M. Tsamparlis \pr {\bf D 24} (1981) 1451; \pl {\bf A 75} (1979) 27.
\bibitem{minkowski} P. Minkowski, \pl {\bf  B 173} (1986) 247.
\bibitem{riess} A.G. Riess et al. \aj {\bf 116} (1998) 1009 .
\bibitem{wang} Y. Wang \apj {\bf 536} (2000) 531.
\bibitem{birk} M. Birkinshaw {\it Phys. Rep.} {\bf 310}(1999) 97.
\bibitem{cavaliere}
A. Cavaliere, R. Fusco--Femiano  \aa {\bf 49} (1976) 137.\\
A. Cavaliere, R. Fusco--Femiano  \aa {\bf 70} (1978) 677.
\bibitem{sarazin} C.L. Sarazin, ``X-Ray Emission from Cluster of Galaxies",
Cambridge Univ. Press, Cambridge (1988).
\bibitem{jetzer}
D. Puy, L. Grenacher, Ph. Jetzer, M. Signore  \aa {\bf 362}
(2000) 415.
\bibitem{cooray98b}
Cooray A., 1998b, A\&A, {\bf 333}, L71.
\bibitem{reese}
Reese E. et al., astro-ph/0205350 (2002).
\end{thebibliography}
\end{document}